\documentclass[prl,twocolumn,superscriptaddress,showpacs,amsmath,amssymb,floatfix]{revtex4}
\usepackage{graphicx,color}% Include figure files
\usepackage{epsfig}

\begin{document}

\title{Enhancement of transmission rates in quantum memory channels with damping}

\author{Giuliano Benenti}
\affiliation{CNISM, CNR-INFM \& Center for Nonlinear and Complex Systems, 
Universit\`a degli Studi dell'Insubria, Via Valleggio 11, 22100 Como, Italy}
\affiliation{Istituto Nazionale di Fisica Nucleare, Sezione di Milano,
via Celoria 16, 20133 Milano, Italy}
\author{Antonio D'Arrigo}
\affiliation{MATIS CNR-INFM, Consiglio Nazionale delle Ricerche and
Dipartimento di Metodologie Fisiche e Chimiche per l'Ingegneria,
Universit\`a di Catania, Viale Andrea Doria 6, 95125 Catania, Italy} 
\author{Giuseppe Falci}
\affiliation{MATIS CNR-INFM, Consiglio Nazionale delle Ricerche and
Dipartimento di Metodologie Fisiche e Chimiche per l'Ingegneria,
Universit\`a di Catania, Viale Andrea Doria 6, 95125 Catania, Italy} 
\date{\today}

% 03.67.Hk Quantum communication
% 89.70.+c Information theory and communication theory 
% 03.65.Yz Decoherence; open systems; quantum statistical methods
% 03.67.Pp Quantum error correction and other methods for protection against decoherence

\begin{abstract}
We consider the transfer of quantum information down a 
single-mode quantum transmission line.  
Such quantum channel is modeled as a damped harmonic oscillator,
the interaction between the information carriers -a train of 
$N$ qubits- and the oscillator being of the Jaynes-Cummings kind. 
Memory effects appear if the state of the oscillator is not reset
after each channel use. We show that the setup 
without resetting is convenient in order to increase the 
transmission rates, both
for the transfer of quantum and classical private information.
Our results can be applied to the micromaser.
\end{abstract}

% 03.67.Hk Quantum communication
% 03.67.-a Quantum information
% 03.65.Yz Decoherence; open systems; quantum statistical methods

\maketitle

Quantum communication channels~\cite{nielsen-chuang,benenti-casati-strini}
use quantum systems as carriers for information. One can employ them 
to transfer classical information, by encoding classical bits by means 
of quantum states. 
Furthermore there are some peculiar issues strictly related to the quantum 
computation: to transfer an (unknown) quantum state 
between different subunits of a quantum computer, to hold in memory a quantum 
state waiting to process it later, to distribute entanglement among different 
parties.

A key problem in quantum information is the determination of the 
classical and quantum \textit{capacities} of noisy quantum channels,
defined as the maximum number of bits/qubits that can be reliably 
transmitted per use of the channel. These quantities characterize the channel,
giving an upper bound to the channel efficiency per use.  

In any realistic implementation, errors occur due to the unavoidable 
coupling of the transmitted quantum systems with an uncontrollable 
environment. Noise can have significant low frequency components, which 
traduce themselves in memory effects, leading to relevant correlations 
in the errors affecting successive transmissions. Important examples 
in this context are photons traveling across fibers with birefringence 
fluctuating with characteristic time scales longer than the separation 
between consecutive light pulses~\cite{banaszek} or low-frequency 
impurity noise in solid state implementations of 
quantum hardware~\cite{solid-state}.
Memory effects become unavoidably relevant when trying to increase
the \textit{transmission rate}~\cite{giovannetti}, that is, to reduce the time interval
that separates two consecutive channel uses.

Quantum channels with memory attracted increasing attention in the 
last years, see~\cite{palma,mancini,giovannetti,werner,datta,njp,virmani,hamada,spins,bosonic,rossini} 
and references therein. Coding theorems have been proved 
for classes of quantum memory channels~\cite{werner,datta}. 
Memory effects have been modeled by 
Markov chains~\cite{palma,mancini,hamada,njp,virmani}
and the quantum capacity has been exactly computed 
for a Markov chain dephasing 
channel~\cite{njp,virmani,hamada}.
Various kind of memory channels are been studied, for example: 
purely dephasing channels~\cite{njp,virmani,rossini},
lossy bosonic channels~\cite{bosonic}; also
spin chains have been studied as 
models for the channel itself~\cite{spins}.
Hamiltonian models of memory 
channels~\cite{njp,virmani,rossini,spins} aim at a 
description directly referring to physical systems and 
enlight another important example of a noisy 
quantum channel, namely the memory of a quantum computer~\cite{barnum}.

In this work the quantum channel is modeled as a damped 
harmonic oscillator, and we
consider transfer of quantum information
through it. A train of $N$ qubits is sent down the channel 
(initially prepared in its ground state) and interacts with it 
during the transit time. 
If the state of the oscillator is not reset after each channel use,
then the action of the channel on the $k$-th qubit 
depends on the previous $k-1$ channel uses. The oscillator acts as
a \textit{local} ``unconventional environment''~\cite{unconventional,mancini,giovannetti,werner}, 
coupled to a memoryless reservoir damping both its phases and populations,  
which mimics any cooling 
process resetting the oscillator to its ground state.
The model is visualized by a qubit-micromaser~\cite{micromaser} system, 
the qubit train being a stream of two-level Rydberg atoms injected at 
low rate into the cavity. Unconventional environments 
capture essentials features of solid state 
circuit-quantum electrodynamics (QED)~\cite{circuitqed} devices and in this context the model may describe 
the architecture of a quantum memory.
The low injection rate is required in order to avoid collective effects such as 
superradiance. Atoms interact with the photon field inside the cavity 
and memory effects are relevant %in high-quality cavities, 
if the lifetime of photons is longer than the time interval 
between two consecutive channel uses. 
In what follows we will show that it is 
convenient to use the channel without 
resetting in order to increase the transmission rates, both
for the transfer of quantum and classical private information.

{\em The quantum capacity.--} 
$N$ channel uses correspond to a $N$-qubit input state
$\rho$, which may be chosen with probability 
$\{p_i\}$ from a given ensemble 
$\{\sigma_i\}$ of the $N$-qubit Liouville space
($\rho=\sum_i p_i \sigma_i$). 
Due to the coupling to uncontrollable degrees of 
freedom, the transmission is in general not fully reliable.
The output is therefore described by a linear, completely positive, 
trace preserving (CPT) map for $N$ uses,  
$\mathcal{E}_N(\rho)$. 
For memoryless channels 
$\mathcal{E}_N=\mathcal{E}_1^{\otimes N}$, where
$\mathcal{E}_1$ indicates the single use, and
the quantum capacity $Q$ can be 
computed as~\cite{lloyd,barnum,devetak}
\begin{equation}
Q \,=\, \lim_{N\to\infty} \frac{Q_N}{N},
\quad \quad 
Q_N\,=\,\max_{\rho} I_c(\mathcal{E}_N,\rho),
\label{qinfo}
\end{equation}
\begin{equation}
I_c(\mathcal{E}_N,\rho)
\,=\,S[\mathcal{E}_N(\rho)]-
S_e(\mathcal{E}_N,\rho).
\label{coinfo}
\end{equation}
Here $S(\rho)=-\mathrm{Tr}[\rho \log_2 \rho]$ is the von 
Neumann entropy, 
$S_e(\mathcal{E}_N,\rho)$ is the \textit{entropy exchange}~\cite{barnum},
defined as 
$S_e(\mathcal{E}_N,\rho)= 
S[(\mathcal{I} \otimes \mathcal{E}_N)(|\psi\rangle\langle \psi|)]$,
where $|\psi\rangle\langle \psi|$ is any purification of $\rho$.
That is, we consider the system ${\textsf S}$, described by the density 
matrix $\rho$, as a part of a larger quantum system ${\textsf R}{\textsf S}$;
$\rho=\mathrm{Tr}_{\textsf R} |\psi\rangle\langle\psi|$ and 
the reference system 
${\textsf R}$ evolves trivially, according to the identity 
superoperator $\mathcal{I}$.
The quantity $I_c(\mathcal{E}_N,\rho)$ is called  
\textit{coherent information}~\cite{barnum}
and must be maximized over over all input states $\rho$.
In general $I_c$ is not subadditive~\cite{barnum}, i.e.
$Q_N/N \ge Q_1$. 
%The equality holds for 
%\textit{degradable channels}~\cite{degradable}, such that 
%if the final state of the environment can be reconstructed from 
%the final state of the system. In this case the 
%regularization $N\to\infty$ in Eq.(\ref{qinfo}) is 
%not necessary and 
%the quantum capacity is given by the ``single-letter'' formula $Q=Q_1$.
When memory effects are taken into account
the channel does not act on each carrier independently, 
${\cal E}_N \neq {\cal E}_1^{\otimes N}$,  
and Eq.~(\ref{qinfo}) in general only provides an upper
bound on the channel capacity. However, for the so-called
\emph{forgetful channels}~\cite{werner}, for which memory effects
decay exponentially with time, a quantum coding theorem 
exists showing
that the upper bound can be saturated~\cite{werner}.

{\em The model.--} 
The overall Hamiltonian governing the dynamics of the 
system ($N$ qubits), a local environment (harmonic oscillator) 
and a reservoir is defined as
($\hbar=1$) 
\begin{eqnarray}
&&\hspace{-0.5cm}{\cal H}(t)={\cal H}_0+V+\delta {\cal H}, \;
                 {\cal H}_0=\nu\left(a^\dagger a+\frac{1}{2}\right) +
                          \frac{\omega}{2}\sum_{k=1}^N \sigma_z^{(k)},
                 \nonumber \\
&&\hspace{1.25cm} V=\lambda \sum_{k=1}^N f_k(t) \left( a^\dagger \sigma_-^{(k)}+
a \sigma_+^{(k)}\right).
\label{model}
\end{eqnarray}
The qubits-oscillator interaction $V$ is of the
Jaynes-Cummings kind, and 
we take $\lambda$ real and positive. 
Coupling is switchable:
$f_k(t)=1$ when qubit $k$ is inside the channel (transit time 
$\tau_p$), $f_k(t)=0$ otherwise. 
The term $\delta {\cal H}$ describes both the reservoir's Hamiltonian
and the local environment-reservoir interaction 
and causes damping of the oscillator
(the cavity mode in the micromaser), that is, 
relaxation and dephasing with  
time scales $\tau_d$ and $\tau_\phi$, respectively. 
Two consecutive qubits entering the channel are separated 
by the time interval $\tau$.

One can argue that the resonant regime $\nu \sim \omega$
is the most significant when describing the coupling 
to modes inducing damping. We work in the
interaction picture, where 
the effective Hamiltonian at resonance is given by
$\tilde{{\cal H}}=
e^{i{\cal H}_0t} (V + \delta {\cal H}) e^{-i{\cal H}_0 t}$
(we will 
omit the tilde from now on).

We assume 
$\tau_p \ll \tau,\tau_\phi,\tau_d$, so that 
non-unitary effects in the evolution of the system and the oscillator
can be ignored during the crossing time $\tau_p$. 
Between two successive pulses the oscillator evolves
according to the standard master equation (obtained after
tracing over the reservoir) 
\begin{equation}
\dot{\rho}_{\textsf c}=
\Gamma \, \left(a \rho_{\textsf c} a^\dagger -
\frac{1}{2} a^\dagger a \rho_{\textsf c} -
\frac{1}{2} \rho_{\textsf c} a^\dagger a \right).
\label{eq:masterdamping}
\end{equation}
The asymptotic decay (channel reset) to the ground state $|0\rangle$ 
takes place with rate $\Gamma$, so that $\tau_d=1/\Gamma$.
We introduce the memory parameter 
$\mu\equiv \tau_d/(\tau+\tau_d)$: fast decay $\tau_d\ll \tau$
yields the 
memoryless limit $\mu\ll 1$, whereas $\mu\lesssim 1$
when memory effects come into play.

{\em The memoryless limit --} In this limit damping acts as 
a built-in reset for the oscillator to its ground state 
$\rho_{\textsf c}(0)=|0\rangle\langle 0|$ after each channel use.
We consider a generic single-qubit input state,
\begin{equation}
\rho_1(0)=
(1-p) |g\rangle\langle g|+
r |g\rangle \langle e| +
r^\star |e\rangle \langle g| +
p |e\rangle \langle e|,
\label{eq:1qubit}
\end{equation}
with $\{|g\rangle,|e\rangle\}$ orthogonal basis for the qubit,
$p$ real and $|r|\le \sqrt{p(1-p)}$.
Given the initial, separable qubit-oscillator state
$\rho_1(0)\otimes \rho_{\textsf c}(0)$, we have
\begin{equation}
\mathcal{E}_1[\rho_1{(0)}]=
{\rm Tr}_{\textsf c}
\{U(\tau_p)[\rho_1(0)\otimes \rho_{\textsf c}(0)]U^\dagger(\tau_p)\},
\label{eq:rho1memoryless}
\end{equation}
with $U(\tau_p)$ unitary time-evolution operator determined by the
undamped Jaynes-Cummings Hamiltonian.
It is easy to obtain~\cite{chen},
in the $\{|g\rangle,|e\rangle\}$ basis,
\begin{equation}
\mathcal{E}_1[\rho_1{(0)}]=
\left[
\begin{array}{cc}
 1-p\cos^2(\lambda\tau_p) & r\cos(\lambda\tau_p)\\
 r^\star\cos(\lambda\tau_p) & p\cos^2(\lambda \tau_p)
\end{array}
\right],
\label{eq:rho1memoryless2}
\end{equation}
with $\lambda$ frequency of the Rabi oscillations between
levels $|e,0\rangle$ and $|g,1\rangle$.
Eq.~(\ref{eq:rho1memoryless2}) corresponds to
an amplitude-damping channel:
$\mathcal{E}_1[\rho_1(0)]= 
\sum_{k=0}^1 E_k \rho_1(0) E_k^\dagger$,
where the Kraus operators~\cite{nielsen-chuang,benenti-casati-strini}
$E_0=|g\rangle \langle g|+
\sqrt{\eta}\, |e\rangle\langle e|$,
$E_1=\sqrt{1-\eta}\, |g\rangle\langle e|$,
with $\eta=\cos^2(\lambda \tau_p)\in [0,1]$. This channel 
is degradable~\cite{giovannettifazio}
and therefore to compute its quantum capacity it is sufficient
to maximize the coherent information over single uses of the channel. 
Maximization is achieved 
by classical states ($r=0$) and one obtains 
$Q=\max_{p\in [0,1]}
\{H_2(\eta p)-H_2[(1-\eta)p]\}$ if $\eta>\frac{1}{2}$,
where $H_2(x)=-x\log_2 x -(1-p) \log_2 (1-x)$ is the binary
Shannon entropy, 
while $Q=0$ when $\eta\le \frac{1}{2}$~\cite{giovannettifazio}. 

{\em Memory channels: validity of Eq.~(\ref{qinfo}) --}
Memory appears in our model when $\tau$ is finite.
To show that the regularized coherent information still
represents the true quantum capacity we follow the arguments made 
for forgetful channels in Ref.~\cite{werner}. 
The key point is the use of a double-blocking strategy mapping,
with a negligible error, the memory channel into a memoryless one.
We consider blocks of $N+L$ uses of the channel and  
do the actual coding and decoding for the first $N$ uses, ignoring the 
remaining $L$ idle uses. We call $\bar{\mathcal{E}}_{N+L}$
the resulting CPT map. 
If we consider $M$ uses of such blocks, the corresponding CPT map
$\bar{\mathcal{E}}_{M(N+L)}$ can be approximated by
the memoryless setting $(\bar{\mathcal{E}}_{(N+L)})^{\otimes M}$.
One can use Eq.~(\ref{qinfo}) to compute the 
quantum capacity if~\cite{werner} 
\begin{equation}
\Vert \bar{\mathcal{E}}_{M(N+L)}(\rho_{\textsf s}) -
(\bar{\mathcal{E}}_{(N+L)})^{\otimes M}(\rho_{\textsf s}) \Vert_1
\leq h\,(M-1) c^{-L},
\label{eq:forgetful}
\end{equation}
where $\rho_{\textsf s}$ is a $M(N+L)$ input state, $h>0$, $c>1$ are 
constant and $\Vert  \rho  \Vert_1={\rm Tr}\sqrt{\rho^\dagger \rho}$ is
the trace norm~\cite{nielsen-chuang}
(note that $c$ and $h$ are independent of the 
input state $\rho_{\textsf s}$).
One can prove~\cite{memorycavitylong} that, due to the exponentially
fast channel (cavity) reset to the ground state,  
inequality (\ref{eq:forgetful}) is fulfilled.
Therefore, quantum capacity can be computed from the maximization
(\ref{qinfo}) of coherent information.

{\em Lower bound for the quantum capacity.--}
For the model Hamiltonian (\ref{model}) 
computation of the coherent information 
for a large number $N$ of channel uses 
is a difficult task, both for analytical and numerical investigations,
even for separable input states, 
$\rho=\rho_1(0)^{\otimes N}$.
Indeed, interaction with the oscillator entangles initially 
independent qubits.
Nevertheless, a lower bound to the quantum capacity can be computed 
if $\tau_\phi\ll \tau$. This happens when additional mechanisms 
of pure dephasing (without relaxation) not explicitly included in 
Eq. (\ref{eq:masterdamping}) dominate the short time dynamics, the
typical situation, e.g., in the solid state. The net effect is that 
the qubit-oscillator phase correlations 
accumulated during their mutual interaction are lost before a new qubit
enters the channel. We account for additional dephasing by 
tracing over each qubit after it
crossed the channel. This enables to 
address the problem
for a very large number of channel uses, at least numerically.
We have $I_c(\mathcal{E}_N,\rho)=\sum_{k=1}^N I_c^{(k)}$, where 
$I_c^{(k)}\equiv I_c[\mathcal{E}_1^{(k)},\rho_1{(0)}]$ and
the CPT map $\mathcal{E}_1^{(k)}$ depends on $k$ due to 
memory effects in the populations of the oscillator.

\begin{figure}
\begin{center}
\epsfxsize=80mm\epsffile{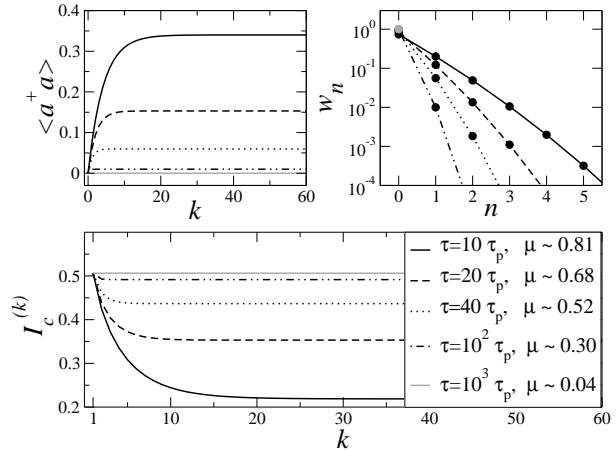}
\caption{Top: $\langle a^\dagger a \rangle$   
as a function of the number $k$ of channel uses (left) and steady-state
populations $w_n$ (right, numerically computed at $k=200$). 
Bottom: coherent information $I_{c}^{(k)}$ as a function of $k$.
Parameter values: $\eta=0.8$ (i.e., $\lambda \tau_p\approx 0.46$) and
$\lambda\tau_d=20$.}
\label{fig:steadystate}
\end{center}
\end{figure}

{\em Steady state --}
In the above strongly dephased regime
the oscillator state $\rho_{\textsf c}^{(k)}$ 
after $k$ channel uses is diagonal and determined by  
the populations $\{w_n^{(k)}\}$. 
The build up of the map that governs the populations 
dynamics requires the computation of the intermediate 
populations $\{\tilde{w}_n^{(k)}\}$, obtained after the 
Jaynes-Cummings interaction
of the $k$-th qubit with the oscillator:
\begin{equation}
\left\{
\begin{array}{l}
\tilde{w}_0^{(k)}=w_0^{(k-1)}[1-p S_1^2]+ w_1^{(k-1)}(1-p)S_1^2,
\\
\tilde{w}_n^{(k)}=w_{n-1}^{(k-1)}pS_n^2
+w_{n}^{(k-1)}[(1-p)C_n^2
\\
\hspace{1cm}+pC_{n+1}^2]
+w_{n+1}^{(k-1)}(1-p)S_{n+1}^2, \; n\ge 1,
\end{array}
\right.
\end{equation}
where we have used the shorthand notation 
$S_n=\sin(\Omega_n\tau_p)$ and 
$C_n=\cos(\Omega_n\tau_p)$, with $\Omega_n=\lambda\sqrt{n}$.
Then the mapping from $\{\tilde{w}_n^{(k)}\}$ to 
$\{{w}_n^{(k)}\}$ is obtained after analytically solving the master 
equation (\ref{eq:masterdamping}) for the populations~\cite{carlo}. 
The overall mapping $\{{w}_n^{(k-1)}\} \to \{{w}_n^{(k)}\}$ is then 
numerically iterated. As shown in Fig.~\ref{fig:steadystate} (top)
a steady state distribution is eventually reached.

Following Eq.~(\ref{coinfo}) we compute 
$I_c^{(k)}=S(\rho_1^{(k)})-S(\rho_{1{\textsf R}}^{(k)})$, with
$\rho_1^{(k)}$ and $\rho_{1{\textsf R}}^{(k)}$
output states for the $k$-th qubit and for the $k$-th qubit plus its
reference system, respectively.
State $\rho_1^{(k)}$ is obtained as in 
Eq.~(\ref{eq:rho1memoryless}), 
but with initial state of the 
oscillator $\rho_{\textsf c}^{(k-1)}$ instead of 
the ground state $\rho_{\textsf c}{(0)}$. We obtain
\begin{equation}
\begin{array}{c}
\rho_1^{(k)}=\sum_{n=0}^\infty w_n^{(k-1)} 
\\
\\
\times
\left[
\begin{array}{cc}
 (1-p)C_n^2+pS_{n+1}^2 & rC_nC_{n+1}\\
 r^\star C_nC_{n+1} & (1-p)S_n^2+pC_{n+1}^2
\end{array}
\right].
\end{array}
\end{equation}
State $\rho_{1{\textsf R}}^{(k)}$ can be conveniently computed
by choosing the purification of $\rho_1(0)$ as in 
Ref.~\cite{giovannettifazio} and again considering the oscillator
initially in the state $\rho_{\textsf c}^{(k-1)}$.
Since $\rho_{\textsf c}^{(k)}$ reaches a steady state, the same
must happen for $I_c^{(k)}$. This expectation is confirmed 
by the numerical data shown in Fig.~\ref{fig:steadystate} (bottom).
The optimization of the regularized coherent information 
(\ref{qinfo}) over separable input states is then simply 
obtained by maximizing the stationary value of the coherent
information over $\rho_1(0)$. The obtained $I_c$-value provides
a lower bound to the quantum capacity of the channel. 

\begin{figure}
\begin{center}
\epsfxsize=80mm\epsffile{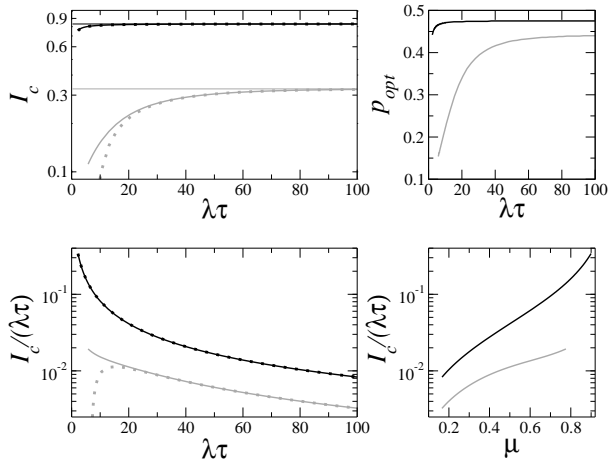}
\caption{Top left: steady state coherent information $I_c$ 
(optimized over separable input states) as a function of 
the dimensionless time separation $\lambda \tau$ 
between consecutive channel uses.
Top right: optimal input state parameter $p_{\rm opt}$ vs. 
$\lambda \tau$.
Bottom left: same data as in the top left panel, but for
the transmission rates $I_c/(\lambda\tau)$. 
Bottom right: transmission rates as a function of the memory
parameter $\mu$. 
Parameter values: 
$\eta=0.95$ ($\lambda \tau_p\approx 0.22$) (black curves),
$\eta=0.7$ ($\lambda \tau_p\approx 0.58$) (gray curves),
$\lambda\tau_d=20$. Dotted curves correspond to 
$p=p_{\rm opt}(\tau\to\infty)$, and show, in the case 
with lower performances of the channel ($\eta=0.7$),
the importance of optimization.}
\label{fig:rates}
\end{center}
\end{figure}

{\em Transmission rates.--} 
The (numerical) optimization is achieved when $r=0$ 
(we have checked it for several values of $\eta$ and $\Gamma$) and $p=p_{\rm opt}$  
in Eq.~(\ref{eq:1qubit}). Note that
$p_{\rm opt}$ may strongly 
depend on the time separation $\tau$ between consecutive channel
uses [see Fig.~\ref{fig:rates} (top right)], namely on the
degree of memory of the channel. 
As shown in Fig.~\ref{fig:rates} (top left) the coherent information,
optimized over separable input states, turns out to be a 
growing function of $\tau$, that is, a 
decreasing function of the degree of memory of the channel.
The memoryless setting $\tau\gg \tau_d$ might appear to be the optimal choice.
However, long waiting times $\tau\gg \tau_d$ are required to reset the
quantum channel (cool the harmonic oscillator/ cavity) to its ground state 
after each channel use, thus reducing the transmission rate.
It appears preferable to consider the quantum transmission rate
$R_Q\equiv Q/\tau$, defined as the maximum number of qubits that 
can be reliably transmitted per unit of time~\cite{giovannetti}. 
Fig.~\ref{fig:rates} (bottom) shows that in order to enhance 
$R_Q$ it is convenient to choose $\tau\ll \tau_d$, namely memory
factors $\mu$ close to 1. In other words, by taking into account memory effects,
one can make more efficient the use of the available transmitting resource.

These results are relevant also for the \textit{secure} transmission of 
classical information, then for cryptographic purposes. 
The reference quantity is, for this case,
the \textit{private classical capacity} $C_p$, defined as 
the capacity for transmitting classical information protected against
an eavesdropper~\cite{devetak}. It was recently shown~\cite{smith} that for 
degradable channels, as it is the case of our model in the memoryless
limit, $C_p=Q$. Since the private classical capacity is always lower bounded 
by the coherent information~\cite{schumacherwestmoreland}, our results 
also show that the setup without resetting is convenient to increase the 
transmission rate $R_p\equiv C_p/\tau$ of private classical information.

{\em Discussion --} Eq.~(\ref{model}) models for instance
dephasing in a micromaser emerging from fluctuations in the
laser field. In the solid state scenario it may describe communication by 
electrons or chiral quasiparticles~\cite{graphene} sent down a mesoscopic channel 
where they interact with optical phonons. As an effective model Eq.~(\ref{model})
has a broad range of applications since the unconventional environment~\cite{unconventional} 
describes the most relevant part of the interaction with a bunch of phonon modes 
producing qubit radiative decay. In such solid-state systems the phonon 
dephasing time $\tau_\phi$ is expected to be much shorter than the phonon 
decay time scale $\tau_d$. In these cases we have shown that a setup without 
memory resetting is convenient in order to increase
the rate of transmission of quantum information and private classical information.
The noisy quantum channel Eq.~(\ref{model})
also describes the dynamics of a quantum memory~\cite{barnum}, which may be 
implemented by coupling $N$ superconducting qubits to a microstrip cavity, 
in a circuit-QED~\cite{circuitqed} architecture. 
In this case, the use of cavities with moderate 
quality factor~\cite{singleph} might be a good trade-off between 
reducing decoherence and avoiding cross-talks generating entanglement
between the qubits crossing the channels. Our results show that in such
situation it is convenient to use the channel without resetting
to increase the rate of sequential processing of each qubit.


\begin{thebibliography}{0}
\bibitem{nielsen-chuang}
M. A. Nielsen and I. L. Chuang,
\textit{Quantum computation and quantum information}
(Cambridge University Press, Cambridge, 2000).

\bibitem{benenti-casati-strini}
%G. Benenti, G. Casati, and G. Strini,
G. Benenti \textit{et al.},
\textit{Principles of quantum computation and information}, vol. II
(World Scientific, Singapore, 2007).

\bibitem{banaszek}
%K. Banaszek, A. Dragan, W. Wasilewski, and C. Radzewicz,
K. Banaszek \textit{et al.}, 
Phys. Rev. Lett. \textbf{92}, 257901 (2004).

\bibitem{solid-state}
%Y. Makhlin, G. Sch\"on, and A. Shnirman, 
Y. Makhlin \textit{et al.}, 
Rev. Mod. Phys. \textbf{73}, 357 (2001);
%E. Paladino, L. Faoro, G. Falci, and R. Fazio, 
E. Paladino \textit{et al.},
Phys. Rev. Lett. \textbf{88}, 228304 (2002);
%G. Falci, A. D'Arrigo, A. Mastellone, and E. Paladino,
G. Falci \textit{et al.},
Phys. Rev. Lett. \textbf{94}, 167002 (2005);
%G. Ithier, E. Collin, P. Joyez, P.J. Meeson, D. Vion, D. Esteve, 
%F. Chiarello, A. Shnirman, Y. Makhlin, J. Schriefl, and G. Sch\"on,
G. Ithier \textit{et al.},
Phys. Rev. B \textbf{72}, 134519 (2005).

\bibitem{giovannetti}
V. Giovannetti, J. Phys. A \textbf{38}, 10989 (2005).

\bibitem{werner}
D. Kretschmann and R. F. Werner, 
Phys. Rev. A \textbf{72}, 062323 (2005).

\bibitem{datta}
N. Datta and T. C. Dorlas, 
J. Phys. A \textbf{40}, 8147 (2007).

\bibitem{palma}
C. Macchiavello and G. M. Palma, 
Phys. Rev. A \textbf{65}, 050301(R) (2002).

\bibitem{mancini}
G. Bowen and S. Mancini, 
Phys. Rev. A \textbf{69}, 012306 (2004). 

\bibitem{hamada}
H. Hamada, J. Math. Phys. \textbf{43} 4382 (2002)

\bibitem{njp}
%A. D'Arrigo, G. Benenti, and G. Falci,
A. D'Arrigo \textit{et al.},
New J. Phys. \textbf{9}, 310 (2007).

\bibitem{virmani}
M. B. Plenio and S. Virmani, 
Phys. Rev. Lett. \textbf{99}, 120504 (2007);
New J. Phys. \textbf{10}, 043032 (2008).

\bibitem{rossini}
D. Rossini \textit{et al.} New J. Phys. \textbf{10} 115009 (2008) 

\bibitem{bosonic}
%O. V. Pilyavets, V. G. Zborovskii, and S.Mancini,
O. V. Pilyavets \textit{et al.}, 
Phys. Rev. A \textbf{77}, 052324 (2008). 

\bibitem{spins}
%A. Bayat, D. Burgarth, S. Mancini, and S. Bose,
A. Bayat \textit{et al.}, 
Phys. Rev. A \textbf{77}, 050306(R) (2008).

\bibitem{barnum}
%H. Barnum, M. A. Nielsen, and B. Schumacher, 
H. Barnum \textit{et al.}, Phys. Rev. A \textbf{57}, 4153 (1998).

\bibitem{unconventional} {\em Focus on Quantum Dissipation in Unconventional Environments},
M. Grifoni and E. Paladino Eds., New J. Phys. {\bf 10} (2008); 
F. Cavaliere et al., New J. Phys. {\bf 10}, 115004 (2008);
A. Garg, Jour. Chem. Phys. {\bf 83}, 4491 (1985); F. Plastina and G. Falci,
Phys. Rev. {\bf B 67}, 224514 (1993).
\bibitem{micromaser}
P. Meystre and M. Sargent III, 
\textit{Elements of quantum optics} (4th Ed.)
(Springer--Verlag, Berlin, 2007).

\bibitem{circuitqed} 
A. Wallraff et al., Nature {\bf 431}, 162 (2004);
J.M. Fink et al., Nature {\bf 454}, 315 (2008).

\bibitem{lloyd}
S. Lloyd, Phys. Rev. A \textbf{55}, 1613 (1997).


\bibitem{devetak}
I. Devetak, IEEE Trans. Inf. Theory \textbf{51}, 44 (2005).

\bibitem{degradable}
I. Devetak and P.W. Shor, Comm. Math. Phys. \textbf{256}, 287 (2005).
%%F. Caruso, V. Giovannetti, and A. S. Holevo,
%F. Caruso \textit{et al.},
%New J. Phys. \textbf{8}, 310 (2006);
%M. W. Wolf and D. P\'erez-Garc\'ia,
%Phys. Rev. A \textbf{75}, 012303 (2007);
%%T. S. Cubitt, M. B. Ruskai, and G. Smith, 
%T. S. Cubitt \textit{et al.},
%J. Math Phys. \textbf{49}, 102104 (2008).

\bibitem{chen}
X.-y. Chen, preprint arXiv:0802.2327 [quant-ph].

\bibitem{giovannettifazio}
V. Giovannetti and R. Fazio, 
Phys. Rev. A \textbf{71}, 032314 (2005).

\bibitem{memorycavitylong}
%A. D'Arrigo, G. Benenti, and G. Falci,
A. D'Arrigo \textit{et al.},
in preparation.

\bibitem{carlo}
%G. G. Carlo, G. Benenti, G. Casati, and C. Mej\'\i a-Monasterio,
G. G. Carlo \textit{et al.},
Phys. Rev. A \textbf{69}, 062317 (2004).

\bibitem{smith}
G. Smith, Phys. Rev. A \textbf{78}, 022306 (2008).

\bibitem{schumacherwestmoreland}
B. W. Schumacher and M. D. Westmoreland, Phys. Rev. Lett. \textbf{80}, 
5695 (1998).

\bibitem{graphene} A. Mitra, I. Aleiner, and A.J. Millis, 
Phys. Rev. B 69, 245302 (2004);
A. Naik \textit{et al.}, Nature {\bf 443}, 193 (2006);
K.S. Novoselov \textit{et al.}, Nature {\bf 438}, 197 (2005). 

\bibitem{singleph} 
A. A. Houck \textit{et al.}, Nature {\bf 449}, 328 (2007). 
\end{thebibliography}
\end{document}